\documentclass[structabstract]{aa}

\usepackage{txfonts}
\usepackage{graphicx}
\usepackage{longtable}
\usepackage{lscape}
\usepackage{latexsym}
\usepackage{url}
\usepackage{natbib}
\bibpunct{(}{)}{;}{a}{}{,}

\begin{document}

\title{Wavelet and $R / S$ analysis of the X-ray flickering of cataclysmic variables}

\author{G.~Anzolin~\inst{1}
\and
F.~Tamburini~\inst{2}
\and
D.~de~Martino~\inst{3}
\and
A.~Bianchini~\inst{2} }

\institute{
ICFO - Institut de Ci\`{e}ncies Fot\`{o}niques, avinguda del Canal Olimpic s/n,
08860 Castelldefels (Barcelona), Spain \\
\email{gabriele.anzolin@icfo.es}
\and
Dipartimento di Astronomia, Universit\`{a} di Padova, vicolo dell'Osservatorio 3,
35122 Padova, Italy \\
\email{[fabrizio.tamburini;antonio.bianchini]@unipd.it}
\and
INAF - Osservatorio Astronomico di Capodimonte, salita Moiariello 16, 80131 Napoli, Italy \\
\email{demartino@oacn.inaf.it}
}

\abstract
{Recently, wavelets and $R / S$ analysis have been used as statistical tools to characterize
the optical flickering of cataclysmic variables.}
{Here we present the first comprehensive study of the statistical properties of X-ray flickering
of cataclysmic variables in order to link them with physical parameters.}
{We analyzed a sample of 97 X-ray light curves of 75 objects of all classes observed with the
\textit{XMM}-Newton space telescope. By using the wavelets analysis, each light curve has
been characterized by two parameters, $\alpha$ and $\Sigma$, that describe the energy
distribution of flickering on different timescales and the strength at a given timescale,
respectively. We also used the $R / S$ analysis to determine the Hurst exponent
of each light curve and define their degree of stochastic memory in time.}
{The X-ray flickering is typically composed of long time scale events
$(1.5 \la \alpha \la 3)$, with very similar strengths in all the subtypes of cataclysmic variables
$(-3 \la \Sigma \la -1.5)$. The X-ray data are distributed in a much smaller area of the
$\alpha -\Sigma$ parameter space with respect to those obtained with optical light curves. The
tendency of the optical flickering in magnetic systems to show higher $\Sigma$ values than the
non-magnetic systems is not encountered in the X-rays. The Hurst exponents estimated for all
light curves of the sample are larger than those found in the visible,
with a peak at 0.82. In particular, we do not obtain values lower than 0.5.
The X-ray flickering presents a persistent memory in time, which seems
to be stronger in objects containing magnetic white dwarf primaries.}
{The similarity of the X-ray flickering in objects of different classes together with the
predominance of a persistent stochastic behavior can be explained it terms of
magnetically-driven accretion processes acting in a considerable fraction of the analyzed
objects.}


\titlerunning{Wavelet and Hurst analysis of the X-ray flickering of CVs}

\maketitle

\section{Introduction}

Cataclysmic variables (CVs) are binary systems in which a late-type secondary star transfers
matter onto a white dwarf (WD) primary. The accretion configuration is different depending on the
magnetic field strength of the WD. In non-magnetic systems the accretion disk extends down to the
WD surface, while a truncated disk may form in moderately magnetic CVs or may not be present at
all in strongly magnetized systems (see \citealt{warner95} for a review).

Owing to the gravitational potential of the compact primary, the accretion of matter produces a
non-negligible flux of X-rays. The mechanism responsible for this emission in non-magnetic CVs
during quiescence is the shock heating that acts in the boundary layer between the accretion disk
and the WD surface \citep{patray85b,patray85a}, while in magnetic systems X-rays are emitted by a
standing shock above the magnetic poles of the WD \citep{aizu73}. In the latter case the X-ray flux
is higher, which makes the magnetic systems the brightest X-ray CVs.

Cataclysmic variables are subdivided into three main classes: novae (classical and recurrent),
dwarf novae (DNe) and nova-like (NL) systems. In addition, there is a class of objects closely
related to novae, the so-called super-soft X-ray sources (SSSs), whose members are characterized by
a prominent soft spectral component due to thermonuclear burning at the WD surface
\citep[see e.g.][]{orio95}. A parallel classification can be made considering the strength of the
magnetic field $B$ of the primary. In this case we have the non-magnetic systems, the intermediate
polars (IPs, $B \sim 5-10$~MG) and the polars ($B \ga 10$~MG). 

The X-ray light curves of CVs may exhibit a variety of modulations. Periodic coherent modulations due
to occultations of the emitting regions can be produced by the orbital motion or by the rotation of
the WD. For magnetic systems periodic modulations can also be ascribed to absorption effects produced
in the magnetically-confined accretion flow [``columns'' in polars \citep{cropper90}, ``curtains''
in IPs \citep{rosetal88}]. Besides this persistent variability, CVs may also show quasi-periodic
oscillations (QPOs) in both the soft and hard X-ray bands as well as completely stochastic brightness
variations, which are usually identified with the term ``flickering'' \citep[see][and references
therein]{kuuetal06}.

Flickering is generally constituted by a sequence of random flares with typical timescales ranging
from a few seconds to a few minutes. This phenomenon has been observed in the X-ray light curves
of CVs of all classes. For example, observations with the \textit{HEAO}-1 satellite allowed the
discovery of soft X-ray flickering in the polar prototype \object{AM~Her} \citep{szketal80}, as well
as hard X-ray aperiodic variability in the DN \object{SS~Cyg} during quiescence \citep{coretal84}.
Flickering at timescales of $\sim 10$~s was detected in the NL \object{TT~Ari} using both
\textit{Einstein} \citep{jenetal83} and \textit{ASCA} \citep{baykiz96} data. Rapid aperiodic
fluctuations were discovered in the light curves of a bunch of CVs also using \textit{ROSAT}
observations \citep{holetal94,roselta95,bucetal98}. A common observational feature is that X-ray
flickering shows a continuous power law frequency spectrum, as it does in the visible region.
A correlation between the time scale of the X-ray flickering activity and that detected in simultaneous
optical observations was found in a number of CVs of different classes, e.g. the IP system
\object{V795~Her} \citep{roselta95}, the polar \object{EF~Eri} \citep{watetal87}, the \object{VY~Scl}
star TT~Ari \citep{jenetal83}, the old nova \object{V603~Aql} \citep{dreetal83} and in the DNe SS~Cyg
and \object{U~Gem} \citep{coretal84}. This observational evidence has been commonly explained as the
result of reprocessing of the X-ray flickering energy.

The properties of flickering have been studied for a long time, especially in the visible region
of the electromagnetic spectrum. However, the origin of optical flickering is still uncertain,
although there is plenty of evidence that it has to be related to the accretion process. It is
very likely that the location of its source should be restricted either to regions very close to
the WD, like the innermost part of the disk, or the hot spot (see \citealt{bruch92} for a thorough
discussion). The connection between X-ray flickering and the accretion onto the WD, instead, is
evident. For highly magnetic CVs, the rapid soft X-ray flares often detected in their light curves
could been explained as the result of a bombardment of the WD surface by random inhomogeneous
structures (``blobs'') present in the accretion streams \citep{kuipri82}. For instance, the features
observed in the light curves of the polar \object{V1309~Ori} \citep{demetal98,schetal05} and of the
asynchronous polar \object{BY~Cam} \citep{ramcro02}, although they are quite peculiar objects,
could be explained with this hypothesis. Also the flares observed in the X-ray light curve of
the old nova \object{GK~Per}, which is an IP, have been interpreted as an indirect indication of
blobby accretion \citep{vrieetal05}. 

In general, it is often possible to obtain some information about the flickering from the
observations of a single object. However, as this is a stochastic process in time, it can be better
studied by means of its statistical properties. \citet[][hereafter FB98]{fribru98} carried out a
statistical analysis of the optical flickering of a large sample of CVs based on the wavelet
analysis \citep{daubechies92} of their light curves. They represented the properties of flickering
in a two-dimensional parameter space and showed that CVs of different classes tend to occupy
different regions of this space. Later \citet[][hereafter TDB09]{tametal09} used the $R / S$
rescaled range analysis \citep{hurst51,huretal65} for the first time as a complementary tool to
characterize the degree of persistence/anti-persistence of the white light flickering of the
IP-class CV \object{V709~Cas}. Motivated by the interesting results presented in these works, we have
therefore used the same statistical techniques to study the X-ray flickering properties of CVs as
a whole as well as a function of their individual classes to link them with physical parameters.

\section{The $\alpha - \Sigma$ parameter space and the Hurst exponent}

A light curve is represented by a set of data points taken at times $t_1, \ldots, t_N$. We assume
that the difference between two consecutive times in the sequence is a constant $\Delta t$, which
usually coincides with the binning time. The calculation of the wavelet transform roughly consists
in decomposing the analyzed signal into a sequence of wavelets all with a general predefined shape
(the mother wavelet), but different positions in time and different scalings. The result is a
two-dimensional set with the coefficients $c_{s,k}$, where $k = 1, \cdots, N$ is the time index and
$s$ is related to the timescale $t_s = 2^s \Delta t$. It is then possible to calculate the quantity
\begin{equation}
S'(t_s) = \frac{2^s}{N} \sum_k c_{s,k}^2 \, ,
\end{equation}
which represents a measure of the variance of the wavelet coefficients on different timescales.
The logarithmic plot of $S'(t_s)$ as a function of $t_s$ is called scalegram \citep{scaetal93}.
\citet{fribru98} introduced a normalized version of the scalegram,
\begin{equation}
S(t_s) = S'(t_s) \frac{N \Delta t}{\sum_{s,k} c_{s,k}^2} \, ,
\end{equation}
and applied it to a large sample of optical flickering light curves of CVs of different classes.
It turned out that almost all scalegrams so obtained were approximately linear functions of
$t_s$, thus permitting a complete representation of the flickering properties with two parameters:
the slope $\alpha$ with respect to the $\log t_s$ axis, which also indicates whether slow or rapid
light fluctuations dominate the stochastic time series, and the strength at a given timescale
$t_{\mathrm{ref}}$, $\Sigma = \log S(t_{\mathrm{ref}})$. Surprisingly, their results indicate that
different subtypes of CVs tend to occupy specific regions of the $\alpha - \Sigma$ parameter space.
This subdivision shows a remarkably small overlap for magnetic NL systems.

The linearity of the scalegrams also reveals that the flickering has the intriguing property
of being self-similar on a wide range of timescales. Self-similarity is a statistical property of
an observable $x(t)$ and is typical of the so-called fractional Brownian motions
\citep{mishura08}. In general, $x(t)$ is said to be self-similar for any scale magnification
$\lambda$ if the following relation holds:
\begin{equation} \label{eqn:selfsim}
x(\lambda t) = \lambda^H x(t) \, .
\end{equation}
The $H$ parameter, called the Hurst exponent, is of fundamental importance in defining the
statistical properties of a physical process. In dynamical systems, where $x(t)$ is a function of
a continuous variable $t$, the Hurst exponent characterizes the stochastic memory in time of a
random process, because it expresses the tendency of the first derivative of $x(t)$ to change sign.
A process is said to be persistent when $H > 1/2$, while an exponent $H < 1/2$ indicates
anti-persistence. In the particular case $H = 1/2$, the process shows a random uncorrelated
behavior with no stochastic memory in time \citep{peters94}.

From an observational point of view, $x(t)$ is represented by a time series with a discrete time
domain. Yet the self-similarity relation expressed by Eq.~\ref{eqn:selfsim} is still valid,
as are the properties of the $H$ parameter. One of the most efficient statistical method used
to estimate the Hurst exponent of a time series is the $R / S$ analysis. The $R / S$ analysis is
non-parametric, in the sense that there are no specific assumptions or requirements for the
distribution of the observables. Moreover, it has been shown to be robust even in the presence of
a discrete noise level \citep{chaetal07}.

Given a set of observables $x(t)$ with $0 \leq t \leq T$, the $R / S$ analysis consists in
subdividing the data set in intervals of length $\tau$ and evaluating the range
\begin{equation}
R(\tau) = \max[X(t,\tau)] - \min[X(t,\tau)]
\end{equation}
and the standard deviation
\begin{equation}
S(\tau) = \sqrt{\frac{\sum_{k=1}^\tau {\left[x(t) - {\left<x\right>}_\tau\right]}^2}{\tau}} \, ,
\end{equation}
where ${\left<x\right>}_\tau$ is the average value of the observable in each sub-interval and
$X(t,\tau ) = \sum_{k=1}^t \left[x(k) - \left<x\right>_\tau \right]$.  The average value of
$\left<R / S\right>_\tau = \left<R(\tau) / S(\tau)\right>$, i.e. a quantity that describes the
``distance'' covered by the observable in units of the local standard deviation, is then
calculated for each sub-interval of equal length $\tau$. For a self-similar process, it
can be shown that
\begin{equation}
\left<R / S\right>_\tau \sim \tau^H \, .
\end{equation}
Hence, the Hurst exponent is represented by the slope of the linear relation between
$\log \tau$ and $\log \left<R / S\right>_\tau$.

It has been also demonstrated that the time-averaged wavelet coefficients depend on the scale
parameter following a power law with exponent $H + 1/2$ \citep{simetal98}. Therefore, the variance
of the wavelet coefficients averaged with respect to the $k$ time index must satisfy the relation
\begin{equation}
{\left<c_{s,k}\right>}_k \sim 2^{\beta s} \, ,
\end{equation}
where $\beta = 2H + 1$ \citep{gaoetal03}. Because of this property, TDB09 proposed to use the
$R / S$ analysis as a complementary tool to characterize the stochastic properties of flickering
by calculating the parameter $\alpha$ through the determination of the Hurst exponent $H$.

\section{Data analysis}

In order to obtain a homogeneous sample of X-ray light curves, we used data collected with the
EPIC cameras (pn, MOS1 and MOS2) onboard the \textit{XMM}-Newton telescope \citep{janetal01},
which provide both good timing accuracy and high sensitivity. The latter property is of particular
importance when dealing with DNe in quiescence, which are typically faint in the X-ray band.
Moreover, thanks to the high spacecraft orbit, \textit{XMM}-Newton is an ideal observatory
to achieve long [up to $\sim 40 \mbox{ hr}$ \citep{baretal99}] uninterrupted observations and
therefore obtain good quality light curves for the $R / S$ analysis.

We used the Catalogue of Cataclysmic Binaries, Low-Mass X-Ray Binaries and Related Objects,
edition 7.11~\footnote{\url{http://www.mpa-garching.mpg.de/RKcat/}} \citep{ritkol03}, to select all
CVs observed with \textit{XMM}-Newton and obtain their J2000 coordinates. We then retrieved all
publicly available data containing processed light curves of these objects from the
\textit{XMM}-Newton Science Archive \citep{arvetal02}. Because the largest part of the observations
were pointed, many of the X-ray sources could be safely identified with the corresponding
object in the catalog. However, a visual check of the automatic identification process was applied for
those data that were not pointed, to exclude spurious sources from our sample. In this way we
found 116 light curves of X-ray sources positively ascribed to CVs. These light curves, as part of the
pipeline products obtained from observations with the EPIC cameras operated either in imaging or timing
mode, were automatically extracted in the 0.2--12~keV energy range, background-subtracted and binned
to obtain a sufficiently good S/N ratio for each bin. We did not apply the heliocentric correction,
because in all cases it is of the order of 0.1~s at most and therefore is negligible compared to
the binning time that is always longer than 10~s.

The quality of the light curves was then checked to select only those suitable for the
statistical analysis. As a first step, although a screening for high background radiation was
already applied to the data during the processing at SSC, we carefully checked the background
status during each observation and discarded all data subjected to high background contamination.
When large gaps were present, the resulting screened light curves were further reduced to have a
continuous coverage of $N > 128$ points and therefore at least six points in the scalegrams. This
also gave reliable results from the $R / S$ analysis \citep{katlhe03}. Furthermore, for some objects
the count rate was too low $(< 0.1 \mbox{ counts s}^{-1})$ and the corresponding light curves were
discarded. After this selection procedure, our final sample constituted a total of 97 light curves
of 75 objects: 23 light curves of 20 DNe, 60 light curves of 46 NL systems, 9 light curves of 5
novae  and 5 light curves of 4 SSSs (see Table~1).

The binning time applied during the pipeline processing at SSC has a lower limit of 10~s for the
brightest objects and is short enough to sufficiently sample the flickering with short timescales.
But this binning time does not allow us to include in the statistical analysis the very rapid
flickering (timescale of $1 - 5 \mbox{ s}$) that might be present in many CVs, especially those
hosting strongly magnetic WDs.

The actual amplitude of the flickering at these very short timescales could be strongly affected
by the shot noise of the EPIC cameras, whose amplitude is roughly proportional to the square root
of the binning time $\Delta t$ assuming a pure Poissonian noise and a constant count rate within
$\Delta t$. As a result, the scalegrams of all light curves would show a flattening at those short
timescales, regardless of the class of the object. With the adopted binning time we expect to avoid
this undesired behavior, which might bias the real statistical properties of flickering.

For each light curve we calculated the $c_{s,k}$ coefficients provided by a discrete wavelet
transform algorithm with a Coiflet C12 mother wavelet and obtained the corresponding scalegrams.
Then, we derived the $\alpha$ parameters by linearly fitting the data points at all time scales,
while $\Sigma$ was calculated at $t_{\mathrm{ref}} = 0.05 \mbox{ hr}$. The wavelet type and the
reference time adopted in our analysis were chosen to allow a direct comparison with the
results of TDB09 and FB98 obtained for the flickering in the visible region. None of the
analyzed scalegrams showed any tendency to be flat at small scales, thus indicating that the S/N
ratio was high enough.

The Hurst exponent was estimated using the $R / S$ analysis and considering all the sub-intervals
of each light curve containing a number of data points in the interval $[10, N]$. The lower limit
was chosen to obtain statistically significant mean values and standard deviations.

\begin{figure}
\centering
\resizebox{8cm}{!}{\includegraphics{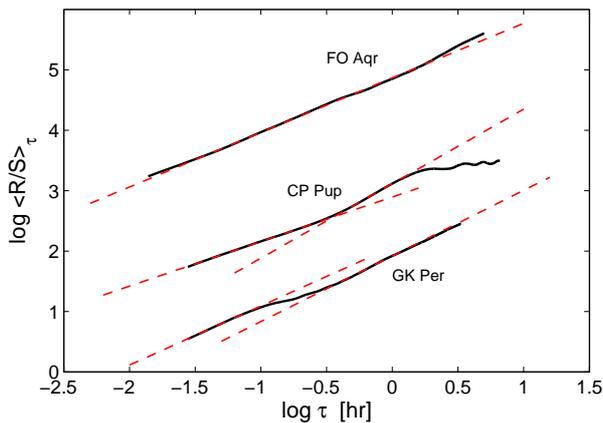}}
\caption{$\log \tau - \log \left<R / S\right>_\tau$ graphs obtained from the $R / S$ analysis of
the X-ray light curves of three CVs of different classes. The curves of CP~Pup and FO~Aqr were shifted
by 1.5 and 3 respectively along the vertical axis. The dashed lines are linear fits of all linear
portions found in each curve.}
\label{fig:rs}
\end{figure}

To obtain the $H$ parameter we had to perform a linear fit of all the data points
appearing in the $\log \tau - \log \left<R / S\right>_\tau$ plane. However, for almost all
light curves we found that there was not just one linear regime of the rescaled range, thus
indicating the presence of different scaling laws at different time scales. Examples of this
behavior are shown in Fig.~\ref{fig:rs} with the data obtained from the $R / S$ analysis of the
X-ray light curves of GK~Per, \object{CP~Pup} and \object{FO~Aqr}. While the curve of the latter
object is almost linear at all scales, for the other two there are two different regimes below and
above $\log (\tau [\mbox{hr}]) \sim -0.5$. An investigation of this peculiar behavior, which is
typical of multifractal systems \citep{stamea88}, is beyond the scope of this paper. Still, the
presence of bends in the scaling law is a well known effect that arises when there are periodic signals.
Indeed, if a periodic modulation with a period $P$ is present in a self-similar time series, then
a break in the initially linear scaling law will appear exactly at $\log P$. Moreover, there will
be more than one break if there are many superposed periodic signals \citep{peters94}. For those data
in the sample that show this phenomenon in the $\log \tau - \log \left<R / S\right>_\tau$ plane, the
breaks are always found at $\log (\tau [\mbox{hr}]) \ga 0.1$ and, therefore, we might expect that
they are likely produced by QPOs or by proper dynamical motions of the CV systems. For instance, the
curve of CP~Pup changes its slope at long time scales (see Fig.~\ref{fig:rs}), where it also shows some
oscillations, because of the appearance of the orbital period $\log (P_\Omega [\mbox{hr}]) = 0.167$
of the system.

The estimation of a unique value of the Hurst exponent is not an adequate way to describe the
self-similarity properties of a light curve because in this way it is impossible to decouple
the different scaling behaviors that might be appearing in different regions of the
$\log \tau - \log \left<R / S\right>_\tau$ plane \citep{simonsen03}. Because we are only interested
in the flickering, we estimated the $H$ exponents considering just the linear portion of
the rescaled range at small scales, i.e. before the appearance of the first break.

The results of our analysis are shown in Table~1, where the entries of the columns are
\begin{enumerate}
\item name of the object;
\item class of the CV (DN = dwarf nova, NL = nova-like, N = nova, SS = super-soft X-ray source);
\item subtype according to \citealt{ritkol03} (AC = AM~CVn, IP = intermediate polar, P = polar,
			SU = SU~UMa, SW = SW~Sex, UG = either U~Gem or SS~Cyg, UX = UX~UMa, VY = VY~Scl, WZ = WZ~Sge,
			ZC = Z~Cam);
\item observation ID number;
\item date of the observation;
\item instrument with which the light curve has been obtained (PN = EPIC-pn, M1 = EPIC-MOS1,
			M2 = EPIC-MOS2);
\item net exposure time;
\item binning time;
\item state of the object at the time of the observation, if known (Q = quiescence, O = outburst,
			H = high-state, L = low state, I = intermediate state, D = decline from nova eruption);
\item mean count rate;
\item $\alpha$ parameter with $1 \sigma$ uncertainty;
\item $\Sigma$ parameter with $1 \sigma$ uncertainty;
\item Hurst exponent with $1 \sigma$ uncertainty.
\end{enumerate}

\begin{figure*}
\centering
\resizebox{17cm}{!}{\includegraphics{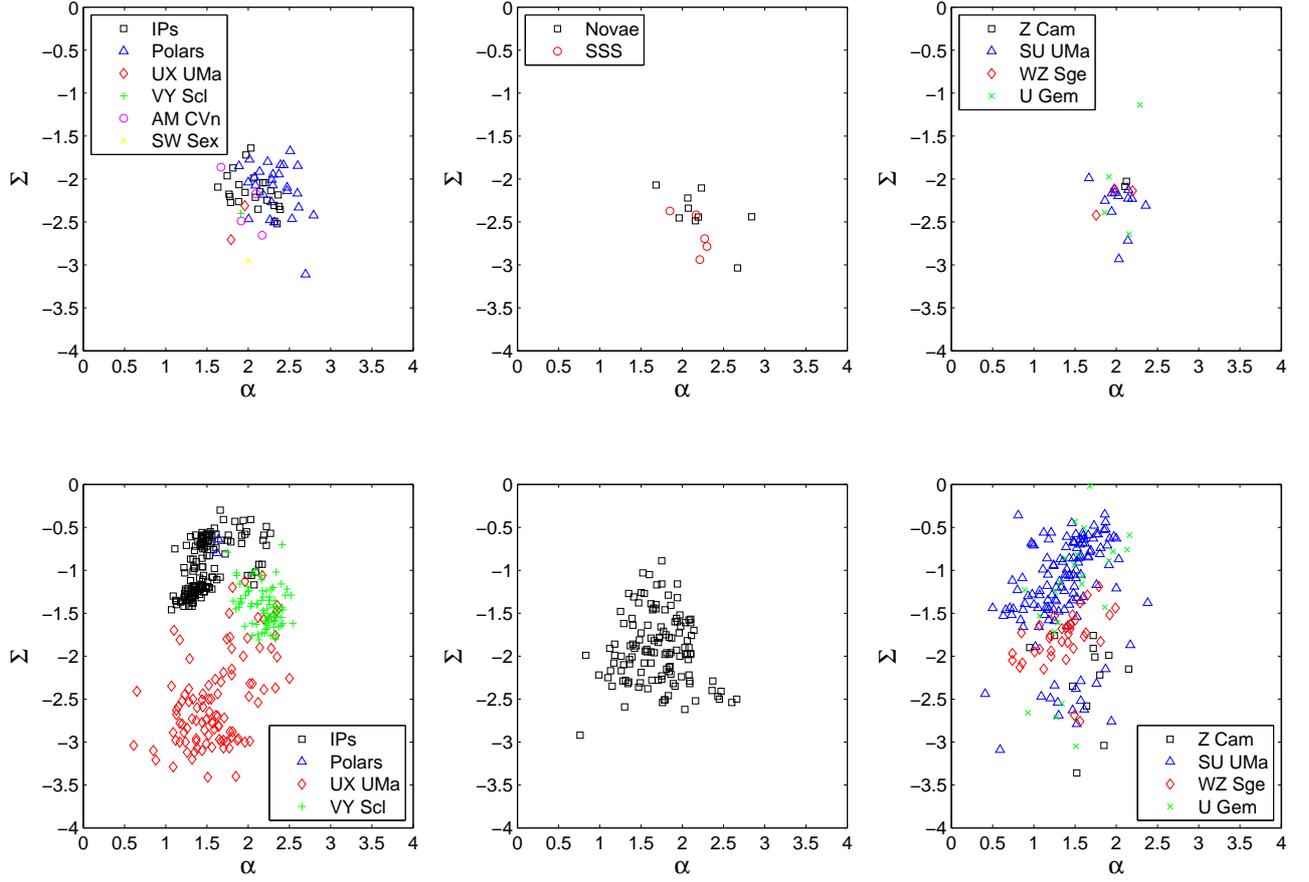}}
\caption{$\alpha-\Sigma$ diagrams of NL systems (left panels), novae and SSSs (central panels) and
DNe (right panels). The diagrams in the top row are obtained from the X-ray data analyzed in this
work, whereas those in the bottom row are obtained from observations in the visible band (FB98).}
\label{fig:alphasigma}
\end{figure*}

\section{Results}

The diagrams representing the distributions of the $\alpha$ and $\Sigma$ parameters obtained with
the wavelet analysis of the X-ray light curves are shown in the upper part of Fig.~\ref{fig:alphasigma}.
For comparison, we also plotted the corresponding diagrams obtained by FB98 from the analysis of
data in the visible band~\footnote{For DNe we only considered visible data obtained during quiescence
because there are just two objects observed in the X-rays during outburst (i.e. \object{WX~Hyi}
and \object{Z~Cha}).}. The data points were subdivided in three diagrams corresponding to the different
classes of CVs and marked with different symbols in function of their subtypes, according to the
classification of \citet{ritkol03}. We also included SSSs, which was not considered in FB98, because
of their affinity with novae and their prominent emission in the soft X-ray domain.

A remarkable result is that all X-ray data points seem to be contained within a small area
delimited by $1.5 \la \alpha \la 3$ and $-3 \la \Sigma \la -1.5$, regardless of the CV class.
If we compare the three X-ray diagrams in Fig.~\ref{fig:alphasigma}, we can see that data points
belonging to objects of different classes totally overlap within this small region. This reveals
a very homogeneous behavior of the X-ray flickering, with just very marginal differences shown in
function of the subtype of CVs. Considering only the X-ray data of NL systems, there is a tendency
for some members of the polar class to have slightly higher $\alpha$ values than IPs and non-magnetic
systems. Moreover, many objects belonging to the NL class show $\Sigma$ values higher than -2,
but novae, SSSs and DNe are always found with $\Sigma \la -2$ (with the noticeable exception of
SS~Cyg). The distribution of the data points of SSSs is very similar to that of novae, although we
cannot exclude that this behavior could be due to the incompleteness of the X-ray sample of objects
belonging to these groups. Dwarf novae appear to be a rather homogeneous class, as was also
pointed out by FB98. Their data points are all located almost in the same region occupied by novae
and SSSs, but with an apparently narrower interval of $\alpha$ values.

\begin{figure}
\centering
\resizebox{8.5cm}{!}{\includegraphics{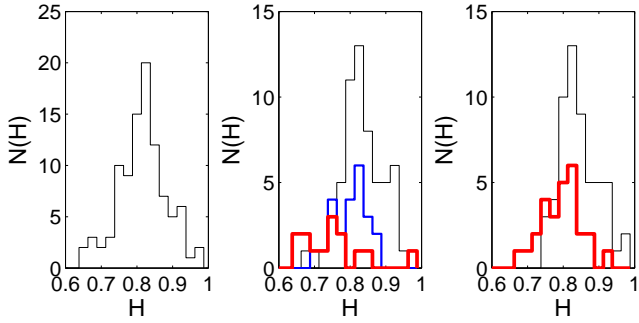}}
\caption{Histograms of the Hurst exponents obtained with the $R / S$ analysis of the X-ray light
curves of CVs. The whole sample is presented in the left panel. The central panel shows the distribution
of $H$ in function of the class of CVs. The histogram marked with a thin line refers to NL systems,
that with an intermediate thickness line refers to DNe and that with a thick line refer to novae and
SSSs. In the right panel we report the distribution of $H$ in function of the magnetic nature of the
WD primary. The histogram marked with a thin line refers to magnetic CVs, while that with a thick line
refers to non-magnetic systems.}
\label{fig:allh}
\end{figure}

Some additional information can be inferred from the results of the $R / S$ analysis. The
distribution of the Hurst exponents of the whole sample of CVs shown in the left panel of
Fig.~\ref{fig:allh} is quite symmetric, with a peak at $H = 0.82$. Moreover, all light curves
present values of $H$ higher than 0.6, with no evidence of purely Gaussian processes $(H = 0.5)$
acting in any of the objects of the sample. The distribution of $H$ in function of the three classes
of CVs (Fig. \ref{fig:allh}, central panel) is strongly biased by the smaller sample of DNe and
novae with respect to the NL systems. We can only say that there is a tendency for the latter to
have higher Hurst exponents with respect to the other two classes. If we instead subdivide the
whole sample into magnetic and non-magnetic systems by checking the additional sub-classifications
reported in the catalog of Ritter \& Kolb, the resulting distributions are more statistically
significant. Then the two distributions (see Fig. \ref{fig:allh}, right panel) are peaked
almost at the same value $H \sim 0.82$, with magnetic systems showing an evident tail towards
higher $H$ and vice-versa for non-magnetic CVs.

\section{Discussion}

\subsection{The visible and X-ray samples} \label{sec:vxsample}

The distribution of the X-ray data points in the $\alpha - \Sigma$ parameter space shows some
important differences from that of the visible data, although a direct comparison between the two
has to be taken with care. Indeed, the numbers of light curves present in the two samples
are very different, about 10 times higher in the FB98 sample than those considered here.
However, it is important to note that many of the points in the visible band diagrams refer to the
same object, while the points obtained in the X-rays generally correspond to only one light curve
per object. Therefore the two samples are comparable in terms of numbers of distinct objects
analyzed (73 in the visible and 75 in the X-rays).

Because many of the data points of the the FB98 sample belong to the same object, their
distribution in the $\alpha - \Sigma$ parameter space also reflects the temporal variation of
the flickering. For this reason, FB98 could analyze the time dependence of the flickering parameter
in some CVs and found a significant variability only for the $\alpha$ parameter.

For the X-ray data, it is impossible to do a similar investigation because none of the objects of the
sample has a comparable number of light curves taken in different epochs. It is only possible to
roughly check the significance of the variations of $\alpha$ and $\Sigma$ parameters by calculating
the absolute value of the difference between their maximum and minimum values ($|\Delta \alpha|$ and
$|\Delta \Sigma|$) and compare them with the corresponding errors.

All objects with at least two X-ray light curves are listed in Table~\ref{tab:timedepX}, together with
the time $\Delta T$ elapsed between the first and the last observation and the resulting $|\Delta \alpha|$
and $|\Delta \Sigma|$ with $1 \sigma$ errors. In general, the variation of the flickering parameters
is not statistically significant when compared with the uncertainties. The only exceptions are for those
CVs that were observed during a transition between two different states, namely the polar \object{CD~Ind}
(from high to low state), the novae \object{V2491~Cyg} and \object{V4743~Sgr} declining to a state of very
low emission, and the DN \object{Z~Cha}, which was in outburst during the first observation. It seems
therefore that the distribution of the X-ray data points of a single CV in the $\alpha - \Sigma$ parameter
space does not strongly depend on the time, at least when the object is observed while it is in the same state.

\begin{table}
\caption{Variation of the flickering parameters in some CVs of the X-ray sample.}
\label{tab:timedepX}
\centering
\begin{tabular}{l r c c}
\hline\hline 
Object name & $\Delta T$ (d) & $|\Delta \alpha|$ & $|\Delta \Sigma|$ \\
\hline
AM Her       &    2.0 & $0.29 \pm 0.22$ & $0.08 \pm 0.39$ \\
CD Ind       &    6.0 & $0.78 \pm 0.32$ & $0.68 \pm 0.22$ \\
EI UMa       &  176.3 & $0.19 \pm 0.11$ & $0.08 \pm 0.20$ \\
EX Hya       &    0.6 & $0.03 \pm 0.18$ & $0.03 \pm 0.29$ \\
HT Cas       &  325.9 & $0.04 \pm 0.14$ & $0.01 \pm 0.21$ \\
MR Vel       &    2.0 & $0.03 \pm 0.17$ & $0.09 \pm 0.28$ \\
NY Lup       &  351.2 & $0.07 \pm 0.14$ & $0.11 \pm 0.25$ \\
OY Car       &   38.4 & $0.36 \pm 0.13$ & $0.21 \pm 0.20$ \\
RXJ0704+2625 &  170.4 & $0.11 \pm 0.33$ & $0.13 \pm 0.54$ \\
RXJ2133+5107 &   38.3 & $0.09 \pm 0.13$ & $0.06 \pm 0.23$ \\
V2400 Oph    &  346.5 & $0.01 \pm 0.13$ & $0.13 \pm 0.22$ \\
V2487 Oph    & 2025.8 & $0.11 \pm 0.17$ & $0.23 \pm 0.34$ \\
V2491 Cyg    &    9.8 & $0.48 \pm 0.14$ & $0.59 \pm 0.24$ \\
V4743 Sgr    &  544.8 & $1.16 \pm 0.14$ & $0.37 \pm 0.23$ \\
XY Ari       &  163.4 & $0.22 \pm 0.17$ & $0.11 \pm 0.27$ \\
Z Cha        &  650.2 & $0.05 \pm 0.13$ & $0.81 \pm 0.21$ \\
\hline
\end{tabular}
\end{table}

\subsection{The X-ray flickering properties}

Taking the issues discussed in Sect.~\ref{sec:vxsample} into consideration, we can see that the
distribution of the X-ray data appears to be much less dispersed than in the visible. The $\alpha$
parameter is found to be positive for all the objects present in the sample, which means that much
of the energy is dissipated in long timescale flickering events. However, we do not find $\alpha$
values lower than $\sim 1.5$ for any CV in the X-ray sample. This differs quite remarkably from the
distribution of the optical data, where some objects show flickering activity with $\alpha \sim 0.5$.
It seems therefore that the X-ray flickering energy is typically dissipated in flares with relatively
longer time scales than those in the optical region. Curiously, there seems to be a kind of cutoff at
$\alpha \sim 3$ in the X-rays and at $\alpha \sim 2.5$ in the visible, which might imply the existence
of an upper limit in the time scale of the flickering independent of the CV class. A more evident
difference is detectable between the distributions of the $\Sigma$ parameter that roughly describes
the strength of the flickering. \citet{fribru98} showed that the energy dissipated in the visible
flickering spans more than 3 orders of magnitude, independently of the CV class, and seems to be
correlated to the subtypes of each class. Instead, the X-ray light curves possess flickering components
with very similar strengths. In particular, the evidence found in the visible data that magnetic NL
systems have higher $\Sigma$ values than the non-magnetic ones is not detectable in the X-ray data.

All subtypes of NL objects populate very similar regions in the X-ray $\alpha - \Sigma$ parameter
space. The tendency of some polars to show slightly higher $\alpha$ values than the other members of
the class suggests that the duration of the X-ray flares might be somewhat correlated to the
modality of accretion, which also depends on the strength of the magnetic field of the WD.

The $R/S$ analysis demonstrates that the X-ray light curves of CVs have, in general, high values
of the Hurst exponent. Therefore, the X-ray flickering has a strong persistent memory in time,
including also extreme cases like GK~Per (in outburst) and \object{RXJ1312+1736} in which $H$ is
close to unity. This is in contrast with the results obtained in the visible band, where TDB09 showed
that the distribution of $H$ has a peak at 0.68 and is strongly asymmetric, with a longer tail
extending towards lower values of $H$. There the light curves show both persistent $(H > 0.5)$
and anti-persistent $(H < 0.5)$ behaviors. Note though that TDB09 calculated the Hurst exponents
from the $\alpha$ parameters of FB98 using the power law relation of the variance of the wavelet
coefficients \citep{gaoetal03}. That is why the derived $H$ values might be affected by the presence
of multiple scaling laws at the different time scales used by FB98 to calculate $\alpha$.

We also found that magnetic CVs tend to posses higher $H$ values with respect to non-magnetic
systems. This implies that the flickering tends to have a more persistent memory in time in those
objects where the accretion is driven by WDs with a magnetic field with significantly detectable
strengths. As proposed by TDB09, this property could be used as a further method to assess the
membership of a CV to a specific class. For instance, we consider \object{LS~Peg}, \object{V426~Oph}
and \object{EI~UMa}, whose classification is still uncertain. The properties of the X-ray
spectra of these objects are similar to those of of IPs \footnote{Note though that EI~UMa is
classified as a DN of the Z~Cam subtype in \citet{ritkol03}.}, but their X-ray light curves do not
show any strong modulation at the spin period of the WD \citep{rametal08}. However, from our analysis
of the X-ray flickering we find $H$ exponents larger than 0.8 in all their light curves (see Table~1),
which could support their magnetic nature.

\subsection{A possible explanation}

It is not straightforward to give an explanation of the difference between the optical and X-ray
flickering properties so far described. In CVs hosting an accretion disk we expect a significant
contribution to the optical flickering due to reprocessing of X-rays \citep{suletal03}, because
the X-ray radiation is most likely produced very close to the WD primary. If these system were also
magnetic, there should be an additional contribution to the optical flickering from the accretion columns.
This might explain the higher $\Sigma$ presented in the visible by magnetic NL systems. However, a
simple reprocessing process cannot explain the observed difference in polar systems, which do not
contain an accretion disk.

\citet{warner04} proposed a scenario in which rapid quasi-coherent brightness modulations in CVs
with an accretion disk could be of magnetic origin. He then argued that a dominant fraction of
apparently non-magnetic systems could host WD primaries with non-negligible fields $(B \la 7 \mbox{ MG})$.
If we adopt this interpretation, we could expect the X-ray flickering to be connected to a
magnetically-driven accretion process in the majority of the objects. This might explain why we
find in the X-ray band a strong similarity among all the classes of CVs. On the other hand, the
appearance of a distinction in the visible between the magnetic and the (supposed) non-magnetic
systems with accretion disk could be explained, apart from reprocessing, if WDs with higher
magnetic fields somehow enhance the optical flickering through processes related to
magneto-hydrodynamical turbulence \citep{kinetal04,dobetal09}.

The link between magnetic accretion and X-ray flickering could explain not only that
magnetic CVs tend to have higher Hurst exponents, but also the evidence that it is, on average,
higher in the X-rays than in the optical region. An argument in favor of this interpretation comes
from studies of artificially generated magnetized plasma, which is a state very similar to that of
the matter within an accretion flow in the proximity of the WD. It has been experimentally found
that edge fluctuations of plasma in several magnetic confinement devices possess a long-range
time correlation, showing values of $H$ between 0.62 and 0.75
\citep[see e.g.][]{caretal99,giletal02}. The same behavior has been noticed in turbulent plasma by
studying the statistics of pulsed phenomena very similar to the blobs supposed to be generating
the X-ray flickering in polars or in IPs with disk-overflow accretion \citep{dencha06,sanetal09}.

\section{Conclusions}

We have used the wavelets and the $R / S$ analysis as statistical tools to characterize the
flickering in a sample of 97 X-ray light curves of CVs of all classes.

In general, the distribution of the X-ray data points in the $\alpha -\Sigma$ parameter space is
almost independent of the class and, therefore, quite different from that in the visible. The
$\alpha$ parameters are all contained in the range $1.5 \la \alpha \la 3$. This implies that the
dissipation of the flickering energy occurs typically in long time scale flares. Moreover, the
strength of the X-ray flickering is found to vary in the region $-3 \la \Sigma \la -1.5$, which is
much smaller than that found in the optical region. There is no evidence for magnetic NL systems
to have higher $\Sigma$ values than the non-magnetic systems. The different behavior in the
visible region and in the X-rays could be explained assuming that part of the optical flickering
in system harboring an accretion disk could be originated by other slower components such as
trailing waves.

All objects in our sample show values of the Hurst exponent higher than 0.5, with a peak of the
distribution at $H = 0.82$. The predominance of a persistent stochastic behavior can be ascribed
to a magnetically-driven accretion process acting in the majority of the CVs considered here.
In particular, the result obtained for magnetic CVs clearly depicts the behavior of a plasma
trapped by the magnetic field in the accretion streams.

However, it must be pointed out that the sample of the X-ray light curves analyzed here is
still not large enough to allow us to draw firm conclusions. Therefore, the analysis of further
data collected in future X-ray observations will permit us to overcome possible selection effects and
to better define the statistical properties of the flickering. On the other hand, a larger sample of
polar systems analyzed in the optical region would allow us to verify our suggestion that the observed
properties of X-ray flickering are intimately related to magnetically-driven accretion processes.

As a final remark, it would have been interesting to use the method proposed by TBD09 to calculate
the Hurst exponent from the $\alpha$ parameter (or vice-versa) and quantitatively demonstrate the impact
of that procedure. However, it was not possible to do it with the X-ray data because for the vast majority
the of the objects in the sample the $R / S$ analysis has been typically stopped at a timescale comparable
to the second or third ``octave'' of the corresponding scalegrams. For this reason, the $H$ values obtained
with the two methods are not comparable because either they refer to different timescales maybe including
multiple Hurst exponents, or the $\alpha$ parameter is obtained by linearly fitting just two or three
points, which is not reliable.

\begin{acknowledgements}
We acknowledge useful discussion about flickering in CVs with B.~Warner and P.~Woudt. GA (partially),
DdM and AB acknowledge financial support from PRIN-INAF under contract PRIN-INAF 2007 N.17. GA
(partially) and DdM acknowledge financial support from ASI/INAF under contract I/023/05/06. FT
acknowledges the financial support from the CARIPARO Foundation inside the 2006 Program of Excellence.
DdM also acknowledges financial support from ASI/INAF under contract I/088/06/0.
\end{acknowledgements}

\bibliographystyle{aa}
\bibliography{xflick}

\longtabL{1}{
\begin{landscape}
\begin{longtable}{lllcclrrccccc}
\caption{Results of the wavelet and $R /S$ analysis of the sample of X-ray light curves (see text for details).} \\
\hline
\hline
Object & Class & Subt. & OBSID & Date obs. & Instr. & Exp. (s) & Bin (s) & State
& Mean rate $(\mbox{cts s}^-{1})$ & $\alpha$ & $\Sigma$ & $H$ \\
\hline
\endfirsthead
\caption{Continued.} \\
\hline
Object & Type & Subt. & OBSID & Date obs. & Instr. & Exp. (s) & Bin (s) & State
& Mean rate $(\mbox{cts s}^-{1})$ & $\alpha$ & $\Sigma$ & $H$ \\
\hline
\endhead
\hline
\endfoot
\hline
\endlastfoot
AB Dra       & DN & ZC & 0111971601 & 2002-10-06 & PN & 10060 &  10 & Q & $ 2.256 \pm 0.014$ & $2.12 \pm 0.06$ & $-2.03 \pm 0.11$ & $0.836 \pm 0.005$ \\
AE Aqr       & NL & IP & 0111180201 & 2001-11-07 & PN & 13930 &  10 & Q & $ 7.253 \pm 0.022$ & $2.39 \pm 0.08$ & $-2.35 \pm 0.14$ & $0.862 \pm 0.008$ \\
AI Tri       & NL &  P & 0306841001 & 2005-08-22 & PN & 19900 &  50 & H & $ 0.466 \pm 0.004$ & $2.32 \pm 0.23$ & $-2.50 \pm 0.34$ & $0.842 \pm 0.008$ \\
AM Her       & NL &  P & 0305240401 & 2005-07-25 & M1 &  9290 &  10 & I & $ 3.782 \pm 0.020$ & $2.47 \pm 0.12$ & $-2.10 \pm 0.21$ & $0.835 \pm 0.005$ \\
             &    &    & 0305240501 & 2005-07-27 & M1 &  7690 &  10 & I & $ 4.145 \pm 0.023$ & $2.18 \pm 0.18$ & $-2.18 \pm 0.32$ & $0.795 \pm 0.003$ \\
AN UMa       & NL &  P & 0109461701 & 2002-05-01 & M1 &  7260 &  20 & H & $ 0.416 \pm 0.007$ & $2.09 \pm 0.13$ & $-2.08 \pm 0.22$ & $0.805 \pm 0.004$ \\
AO Psc       & NL & IP & 0009650101 & 2001-06-09 & M1 & 40830 &  10 & Q & $ 1.784 \pm 0.006$ & $2.07 \pm 0.23$ & $-1.99 \pm 0.40$ & $0.852 \pm 0.009$ \\
BY Cam       & NL &  P & 0109460901 & 2001-08-26 & M1 &  5770 &  10 & H & $ 1.534 \pm 0.016$ & $2.08 \pm 0.16$ & $-1.98 \pm 0.29$ & $0.739 \pm 0.003$ \\
CAL83        & SS &    & 0123510101 & 2000-04-23 & PN & 10020 &  10 &   & $ 4.363 \pm 0.020$ & $2.17 \pm 0.16$ & $-2.42 \pm 0.29$ & $0.679 \pm 0.003$ \\
CAL87        & SS &    & 0153250101 & 2003-04-18 & PN & 76760 &  20 &   & $ 1.721 \pm 0.004$ & $2.21 \pm 0.19$ & $-2.94 \pm 0.29$ & $0.660 \pm 0.008$ \\
CD Ind       & NL &  P & 0111250101 & 2002-03-27 & PN & 13260 &  10 & H & $ 3.946 \pm 0.017$ & $2.30 \pm 0.10$ & $-2.07 \pm 0.17$ & $0.815 \pm 0.002$ \\
             &    &    & 0111251401 & 2002-03-28 & PN &  8650 &  10 & H & $ 3.194 \pm 0.019$ & $2.28 \pm 0.15$ & $-2.01 \pm 0.26$ & $0.824 \pm 0.002$ \\
             &    &    & 0111251501 & 2002-03-29 & PN & 13260 &  10 & I & $ 2.138 \pm 0.012$ & $2.14 \pm 0.08$ & $-1.91 \pm 0.13$ & $0.823 \pm 0.003$ \\
             &    &    & 0111251601 & 2002-03-30 & PN & 13250 &  10 & I & $ 2.815 \pm 0.014$ & $2.02 \pm 0.12$ & $-1.77 \pm 0.21$ & $0.910 \pm 0.002$ \\
             &    &    & 0111251801 & 2002-03-31 & PN & 14020 &  10 & I & $ 2.048 \pm 0.012$ & $2.24 \pm 0.03$ & $-1.80 \pm 0.06$ & $0.859 \pm 0.002$ \\
             &    &    & 0111251901 & 2002-04-01 & PN & 14070 &  30 & I & $ 0.502 \pm 0.005$ & $2.79 \pm 0.29$ & $-2.42 \pm 0.46$ & $0.822 \pm 0.003$ \\
             &    &    & 0111251701 & 2002-04-02 & PN & 14320 &  40 & I & $ 0.470 \pm 0.005$ & $2.26 \pm 0.14$ & $-2.48 \pm 0.21$ & $0.833 \pm 0.004$ \\
CE Gru       & NL &  P & 0109463501 & 2001-10-31 & PN &  7616 &  40 & H & $ 0.496 \pm 0.010$ & $2.53 \pm 0.20$ & $-2.47 \pm 0.34$ & $0.792 \pm 0.008$ \\
CP Pup       &  N &    & 0304010101 & 2005-06-04 & PN & 49960 &  20 & Q & $ 1.099 \pm 0.004$ & $2.16 \pm 0.12$ & $-2.49 \pm 0.19$ & $0.747 \pm 0.003$ \\
CR Boo       & DN & SU & 0202890201 & 2003-12-28 & PN & 20100 &  30 & Q & $ 0.730 \pm 0.006$ & $2.14 \pm 0.20$ & $-2.72 \pm 0.30$ & $0.701 \pm 0.007$ \\
DW UMa       & NL & SW & 0142970101 & 2003-10-18 & PN & 24000 & 160 &   & $ 0.104 \pm 0.002$ & $2.00 \pm 0.55$ & $-2.95 \pm 0.76$ & $0.776 \pm 0.007$ \\
EI UMa       & NL & IP & 0111971301 & 2002-05-10 & PN &  4700 &  10 & Q & $ 6.191 \pm 0.036$ & $2.27 \pm 0.07$ & $-2.14 \pm 0.13$ & $0.847 \pm 0.001$ \\
             &    &    & 0111971701 & 2002-11-02 & PN &  8270 &  10 & I & $ 6.368 \pm 0.027$ & $2.09 \pm 0.08$ & $-2.21 \pm 0.14$ & $0.865 \pm 0.002$ \\
EP Dra       & NL &  P & 0109464501 & 2002-10-18 & PN & 17600 &  20 & H & $ 0.449 \pm 0.005$ & $2.43 \pm 0.10$ & $-1.84 \pm 0.17$ & $0.849 \pm 0.004$ \\
EX Hya       & NL & IP & 0111020101 & 2000-07-01 & M1 & 43590 &  10 & Q & $ 6.017 \pm 0.011$ & $2.35 \pm 0.12$ & $-2.52 \pm 0.21$ & $0.804 \pm 0.001$ \\
             &    &    & 0111020201 & 2000-07-01 & M1 & 35060 &  10 & Q & $ 6.075 \pm 0.013$ & $2.32 \pm 0.13$ & $-2.49 \pm 0.21$ & $0.792 \pm 0.001$ \\
EV UMa       & NL &  P & 0109462201 & 2001-12-08 & PN &  5040 &  20 & H & $ 1.714 \pm 0.018$ & $2.61 \pm 0.30$ & $-2.33 \pm 0.52$ & $0.883 \pm 0.010$ \\
FO Aqr       & NL & IP & 0009650201 & 2001-05-12 & M1 & 36010 &  10 & H & $ 0.664 \pm 0.004$ & $2.03 \pm 0.06$ & $-1.64 \pm 0.10$ & $0.793 \pm 0.002$ \\
GD552        & DN & WZ & 0301830301 & 2005-06-18 & M1 & 27300 &  20 & Q & $ 0.307 \pm 0.003$ & $1.98 \pm 0.09$ & $-2.12 \pm 0.14$ & $0.713 \pm 0.003$ \\
GG Leo       & NL &  P & 0109461401 & 2002-05-13 & M1 &  7840 &  20 & H & $ 0.391 \pm 0.007$ & $2.47 \pm 0.27$ & $-2.14 \pm 0.46$ & $0.796 \pm 0.006$ \\
GK Per       &  N & IP & 0154550101 & 2002-03-09 & PN & 29520 &  10 & O & $10.745 \pm 0.019$ & $2.23 \pm 0.15$ & $-2.11 \pm 0.25$ & $0.985 \pm 0.002$ \\
GP Com       & NL & AC & 0017940101 & 2001-01-03 & PN & 51270 &  10 & L & $ 2.040 \pm 0.006$ & $2.09 \pm 0.07$ & $-2.17 \pm 0.11$ & $0.769 \pm 0.003$ \\
HP Lib       & NL & AC & 0202890101 & 2004-01-28 & PN & 20100 &  30 & H & $ 0.553 \pm 0.005$ & $1.92 \pm 0.19$ & $-2.49 \pm 0.30$ & $0.722 \pm 0.002$ \\
HS Cam       & NL &  P & 0143430101 & 2003-10-13 & PN & 14650 &  10 &   & $ 1.879 \pm 0.011$ & $2.51 \pm 0.12$ & $-1.67 \pm 0.20$ & $0.875 \pm 0.003$ \\
HT Cam       & NL & IP & 0144840101 & 2003-03-24 & M1 & 40480 &  20 & Q & $ 0.412 \pm 0.003$ & $1.77 \pm 0.18$ & $-2.18 \pm 0.28$ & $0.788 \pm 0.006$ \\
HT Cas       & DN & SU & 0111310101 & 2002-08-20 & PN & 47980 &  20 & Q & $ 0.937 \pm 0.004$ & $1.95 \pm 0.06$ & $-2.17 \pm 0.09$ & $0.834 \pm 0.004$ \\
             &    &    & 0152490201 & 2003-07-12 & PN & 42600 &  30 & Q & $ 0.722 \pm 0.004$ & $1.98 \pm 0.13$ & $-2.16 \pm 0.19$ & $0.846 \pm 0.007$ \\
HU Aqr       & NL &  P & 0110860101 & 2002-05-16 & M1 & 37658 &  30 & I & $ 0.142 \pm 0.002$ & $1.89 \pm 0.16$ & $-1.85 \pm 0.26$ & $0.934 \pm 0.012$ \\
IX Vel       & NL & UX & 0111971001 & 2001-12-06 & PN & 19020 &  10 & H & $ 1.919 \pm 0.010$ & $1.96 \pm 0.14$ & $-2.31 \pm 0.24$ & $0.674 \pm 0.004$ \\
LS Peg       & NL & IP & 0306290201 & 2005-06-08 & PN & 42600 &  30 &   & $ 0.603 \pm 0.003$ & $1.88 \pm 0.13$ & $-2.26 \pm 0.20$ & $0.813 \pm 0.004$ \\
MR Vel       & SS &    & 0111150101 & 2000-12-16 & PN & 56220 &  10 & Q & $ 5.423 \pm 0.009$ & $2.27 \pm 0.11$ & $-2.70 \pm 0.18$ & $0.716 \pm 0.004$ \\
             &    &    & 0111150201 & 2000-12-18 & PN & 56820 &  10 & Q & $ 6.076 \pm 0.010$ & $2.30 \pm 0.13$ & $-2.79 \pm 0.21$ & $0.699 \pm 0.005$ \\ 
MU Cam       & NL & IP & 0306550101 & 2006-04-06 & M2 & 28160 &  20 &   & $ 0.395 \pm 0.003$ & $1.82 \pm 0.15$ & $-1.87 \pm 0.23$ & $0.810 \pm 0.003$ \\
NY Lup       & NL & IP & 0105460301 & 2000-09-07 & PN & 19420 &  10 &   & $ 5.788 \pm 0.017$ & $2.14 \pm 0.11$ & $-2.15 \pm 0.19$ & $0.834 \pm 0.004$ \\
             &    &    & 0105460501 & 2001-08-24 & PN & 15130 &  10 &   & $ 8.604 \pm 0.023$ & $2.21 \pm 0.10$ & $-2.04 \pm 0.17$ & $0.855 \pm 0.004$ \\
OY Car       & DN & SU & 0099020301 & 2000-06-29 & PN & 51320 &  20 & Q & $ 1.109 \pm 0.004$ & $2.03 \pm 0.06$ & $-2.20 \pm 0.09$ & $0.830 \pm 0.004$ \\
             &    &    & 0128320301 & 2000-08-07 & M1 & 14250 &  30 & Q & $ 0.185 \pm 0.003$ & $1.67 \pm 0.12$ & $-1.99 \pm 0.18$ & $0.789 \pm 0.005$ \\
PQ Gem       & NL & IP & 0109510301 & 2002-10-07 & M1 & 35780 &  10 &   & $ 1.564 \pm 0.006$ & $1.89 \pm 0.10$ & $-2.07 \pm 0.16$ & $0.831 \pm 0.006$ \\
QS Tel       & NL &  P & 0404710401 & 2006-09-30 & PN & 21222 & 100 & I & $ 0.188 \pm 0.003$ & $2.69 \pm 0.14$ & $-3.11 \pm 0.20$ & $0.939 \pm 0.015$ \\
RXJ0425-5714 & NL &  P & 0148000601 & 2003-09-16 & PN & 12220 &  10 &   & $ 2.700 \pm 0.014$ & $2.39 \pm 0.10$ & $-1.84 \pm 0.17$ & $0.930 \pm 0.004$ \\
RXJ0439-6809 & SS &    & 0008020101 & 2001-11-01 & PN & 19500 &  60 & Q & $ 0.361 \pm 0.004$ & $1.85 \pm 0.26$ & $-2.37 \pm 0.39$ & $0.667 \pm 0.023$ \\
RXJ0704+2625 & NL & IP & 0401650101 & 2006-10-04 & PN & 10240 &  20 &   & $ 1.156 \pm 0.010$ & $1.63 \pm 0.24$ & $-2.10 \pm 0.40$ & $0.761 \pm 0.011$ \\
             &    &    & 0401650301 & 2007-03-23 & PN &  8060 &  20 &   & $ 1.285 \pm 0.012$ & $1.75 \pm 0.22$ & $-1.96 \pm 0.37$ & $0.766 \pm 0.004$ \\
RXJ1007-2017 & NL &  P & 0109461301 & 2001-12-07 & PN &  4760 &  10 & H & $ 4.328 \pm 0.030$ & $2.37 \pm 0.08$ & $-1.94 \pm 0.15$ & $0.922 \pm 0.003$ \\
RXJ1050-1404 & DN & WZ & 0301830101 & 2005-06-16 & PN & 25200 & 100 &   & $ 0.213 \pm 0.002$ & $1.76 \pm 0.23$ & $-2.42 \pm 0.32$ & $0.816 \pm 0.003$ \\
RXJ1312+1736 & NL &  P & 0200000101 & 2004-06-28 & PN & 31677 & 100 &   & $ 0.201 \pm 0.003$ & $2.00 \pm 0.28$ & $-2.47 \pm 0.41$ & $0.982 \pm 0.005$ \\
RXJ1730-0559 & NL & IP & 0302100201 & 2005-08-29 & PN & 11290 &  10 &   & $ 4.175 \pm 0.019$ & $1.97 \pm 0.13$ & $-1.72 \pm 0.22$ & $0.929 \pm 0.004$ \\
RXJ1803+4012 & NL & IP & 0501230101 & 2007-08-31 & PN & 20120 &  40 &   & $ 0.560 \pm 0.005$ & $1.79 \pm 0.17$ & $-2.27 \pm 0.25$ & $0.832 \pm 0.010$ \\
RXJ2133+5107 & NL & IP & 0302100101 & 2005-05-29 & PN & 13600 &  10 &   & $ 5.112 \pm 0.019$ & $2.31 \pm 0.11$ & $-2.31 \pm 0.19$ & $0.828 \pm 0.004$ \\
             &    &    & 0302100301 & 2005-07-06 & PN &  9890 &  10 &   & $ 4.794 \pm 0.022$ & $2.23 \pm 0.07$ & $-2.25 \pm 0.13$ & $0.835 \pm 0.006$ \\
SS Aur       & DN & UG & 0502640201 & 2008-04-07 & PN & 36150 &  30 &   & $ 0.702 \pm 0.004$ & $2.16 \pm 0.14$ & $-2.64 \pm 0.22$ & $0.780 \pm 0.006$ \\
SS Cyg       & DN & UG & 0111310201 & 2001-06-05 & PN & 11870 &  10 & Q & $42.503 \pm 0.059$ & $2.28 \pm 0.06$ & $-1.14 \pm 0.14$ & $0.843 \pm 0.004$ \\
SU UMa       & DN & SU & 0111970801 & 2002-05-05 & PN & 11670 &  10 & Q & $ 6.071 \pm 0.022$ & $2.36 \pm 0.11$ & $-2.31 \pm 0.18$ & $0.863 \pm 0.003$ \\
T Leo        & DN & SU & 0111970701 & 2002-06-01 & PN & 10020 &  10 & Q & $ 3.583 \pm 0.018$ & $2.19 \pm 0.10$ & $-2.23 \pm 0.18$ & $0.802 \pm 0.004$ \\
TY Psa       & DN & SU & 0111970101 & 2001-11-28 & PN & 10040 &  20 & Q & $ 1.270 \pm 0.011$ & $1.94 \pm 0.13$ & $-2.38 \pm 0.22$ & $0.806 \pm 0.004$ \\
U Gem        & DN & UG & 0110070401 & 2002-04-13 & M1 & 22430 &  10 & Q & $ 0.873 \pm 0.006$ & $1.91 \pm 0.09$ & $-1.97 \pm 0.16$ & $0.738 \pm 0.007$ \\
UU Col       & NL & IP & 0201290201 & 2004-08-21 & PN & 26100 &  30 &   & $ 0.635 \pm 0.004$ & $1.78 \pm 0.11$ & $-2.21 \pm 0.18$ & $0.910 \pm 0.003$ \\
UX UMa       & NL & UX & 0084190201 & 2002-06-12 & PN & 46100 & 100 & H & $ 0.182 \pm 0.001$ & $1.79 \pm 0.17$ & $-2.71 \pm 0.24$ & $0.794 \pm 0.012$ \\
UZ For       & NL &  P & 0111320101 & 2002-08-08 & M1 & 29280 &  60 & L & $ 0.072 \pm 0.001$ & $2.28 \pm 0.14$ & $-2.27 \pm 0.21$ & $0.852 \pm 0.005$ \\
V347 Pav     & NL &  P & 0109462901 & 2002-03-16 & M1 &  5840 &  20 & H & $ 0.305 \pm 0.007$ & $2.60 \pm 0.26$ & $-2.17 \pm 0.44$ & $0.898 \pm 0.005$ \\
V396 Hya     & NL & AC & 0302160201 & 2005-07-20 & PN & 28000 &  40 &   & $ 0.469 \pm 0.004$ & $2.17 \pm 0.12$ & $-2.66 \pm 0.18$ & $0.792 \pm 0.008$ \\
V407 Vul     & NL & AC & 0109500201 & 2003-11-15 & M2 &  8789 &  30 &   & $ 0.180 \pm 0.005$ & $1.67 \pm 0.32$ & $-1.86 \pm 0.53$ & $0.936 \pm 0.031$ \\
V426 Oph     & DN & ZC & 0306290101 & 2006-03-04 & M1 & 36190 &  10 &   & $ 2.284 \pm 0.007$ & $2.11 \pm 0.04$ & $-2.09 \pm 0.08$ & $0.842 \pm 0.003$ \\
V442 Oph     & NL & VY & 0305440801 & 2005-09-28 & PN & 11640 &  60 & H & $ 0.342 \pm 0.005$ & $1.91 \pm 0.17$ & $-2.40 \pm 0.27$ & $0.934 \pm 0.003$ \\
V834 Cen     & NL &  P & 0405520301 & 2007-01-30 & M1 & 44680 &  10 & L & $ 1.166 \pm 0.005$ & $2.00 \pm 0.09$ & $-2.04 \pm 0.15$ & $0.738 \pm 0.008$ \\
V1223 Sgr    & NL & IP & 0145050101 & 2003-04-13 & M1 & 38680 &  10 &   & $ 3.844 \pm 0.009$ & $2.12 \pm 0.09$ & $-2.35 \pm 0.15$ & $0.770 \pm 0.001$ \\
V1309 Ori    & NL &  P & 0010620101 & 2001-03-18 & PN & 26580 &  30 & H & $ 0.804 \pm 0.005$ & $2.30 \pm 0.08$ & $-1.95 \pm 0.13$ & $0.771 \pm 0.004$ \\
V2301 Oph    & NL &  P & 0109465301 & 2004-09-06 & PN & 16680 &  10 & H & $ 3.577 \pm 0.014$ & $2.60 \pm 0.07$ & $-1.85 \pm 0.13$ & $0.879 \pm 0.003$ \\
V2400 Oph    & NL & IP & 0105460101 & 2000-09-18 & PN &  7820 &  10 &   & $ 6.361 \pm 0.028$ & $2.36 \pm 0.05$ & $-2.19 \pm 0.09$ & $0.885 \pm 0.006$ \\
             &    &    & 0105460601 & 2001-08-30 & PN & 15620 &  10 &   & $ 8.359 \pm 0.023$ & $2.37 \pm 0.12$ & $-2.32 \pm 0.20$ & $0.899 \pm 0.008$ \\
V2487 Oph    &  N &    & 0085581401 & 2001-09-05 & PN &  5040 &  20 & D & $ 1.049 \pm 0.014$ & $2.07 \pm 0.10$ & $-2.34 \pm 0.18$ & $0.775 \pm 0.003$ \\
             &    &    & 0085582001 & 2002-09-24 & PN &  6600 &  20 & D & $ 1.046 \pm 0.012$ & $2.07 \pm 0.16$ & $-2.22 \pm 0.26$ & $0.742 \pm 0.005$ \\
             &    &    & 0401660101 & 2007-03-24 & PN & 33380 &  20 & D & $ 1.242 \pm 0.006$ & $1.96 \pm 0.14$ & $-2.45 \pm 0.22$ & $0.815 \pm 0.003$ \\
V2491 Cyg    &  N &    & 0552270501 & 2008-05-20 & M1 & 39030 &  10 & D & $33.912 \pm 0.029$ & $2.67 \pm 0.10$ & $-3.04 \pm 0.17$ & $0.782 \pm 0.004$ \\
             &    &    & 0552270601 & 2008-05-30 & M1 & 29920 &  10 & D & $ 9.378 \pm 0.017$ & $2.19 \pm 0.11$ & $-2.44 \pm 0.18$ & $0.839 \pm 0.003$ \\
V4743 Sgr    &  N &    & 0127720501 & 2003-04-04 & M2 &  2900 &  10 & D & $22.853 \pm 0.089$ & $2.84 \pm 0.10$ & $-2.44 \pm 0.19$ & $0.649 \pm 0.011$ \\
             &    &    & 0204690101 & 2004-09-30 & M1 & 22240 &  40 & D & $ 0.111 \pm 0.002$ & $1.68 \pm 0.09$ & $-2.07 \pm 0.13$ & $0.744 \pm 0.008$ \\
VW Hyi       & DN & SU & 0111970301 & 2001-10-19 & PN & 16130 &  10 & Q & $ 2.246 \pm 0.011$ & $2.14 \pm 0.14$ & $-2.23 \pm 0.24$ & $0.755 \pm 0.010$ \\
VZ Sex       & DN & UG & 0201290301 & 2004-05-18 & PN & 35790 &  30 &   & $ 0.726 \pm 0.004$ & $1.86 \pm 0.13$ & $-2.39 \pm 0.19$ & $0.768 \pm 0.004$ \\
WW Cet       & DN & IP & 0111970901 & 2001-12-06 & PN &  9520 &  10 & Q & $ 4.208 \pm 0.021$ & $2.30 \pm 0.07$ & $-2.27 \pm 0.13$ & $0.806 \pm 0.005$ \\
WX Hyi       & DN & SU & 0111970401 & 2002-01-08 & PN &  8700 &  30 & O & $ 0.769 \pm 0.009$ & $1.86 \pm 0.16$ & $-2.25 \pm 0.25$ & $0.879 \pm 0.005$ \\
WZ Sge       & DN & WZ & 0150100101 & 2003-05-16 & PN &  8070 &  10 & Q & $ 2.835 \pm 0.018$ & $2.20 \pm 0.10$ & $-2.14 \pm 0.18$ & $0.751 \pm 0.008$ \\
XY Ari       & NL & IP & 0110660101 & 2000-08-26 & PN & 17340 &  20 & Q & $ 0.877 \pm 0.007$ & $2.18 \pm 0.14$ & $-2.05 \pm 0.22$ & $0.833 \pm 0.004$ \\
             &    &    & 0112510301 & 2001-02-05 & PN & 26700 &  30 & Q & $ 0.786 \pm 0.005$ & $1.96 \pm 0.11$ & $-2.16 \pm 0.16$ & $0.894 \pm 0.004$ \\
YZ Cnc       & DN & SU & 0152530101 & 2002-10-05 & PN & 35060 &  10 & Q & $ 2.887 \pm 0.009$ & $2.15 \pm 0.04$ & $-2.13 \pm 0.06$ & $0.824 \pm 0.001$ \\
Z Cha        & DN & SU & 0205770101 & 2003-12-19 & PN & 99580 &  20 & O & $ 1.020 \pm 0.003$ & $1.98 \pm 0.10$ & $-2.12 \pm 0.15$ & $0.825 \pm 0.003$ \\
             &    &    & 0306560301 & 2005-09-30 & PN & 87180 &  60 & Q & $ 0.302 \pm 0.001$ & $2.03 \pm 0.10$ & $-2.93 \pm 0.14$ & $0.747 \pm 0.003$ \\
\end{longtable}
\end{landscape}}

\end{document}